\documentclass[%
reprint,
superscriptaddress,
amsmath,amssymb,
aps,
prab,
floatfix,
]{revtex4-2}

\usepackage{graphicx}% Include figure files
\usepackage{dcolumn}% Align table columns on decimal point
\usepackage{bm}% bold math
\usepackage{xcolor}
\begin{document}
	
	\newcommand{\I}{\mathrm{i}}
	\newcommand{\E}{\mathrm{e}}
	\newcommand{\D}{\,\mathrm{d}}
	
	\title{%PROPOSAL: %roposal Part II:  Utilization of 
 X-ray Free-Electron Laser accelerator to study on Quantum Electro-Dynamics} 

	\author{Svitozar Serkez}
    \email{svitozar.serkez@xfel.eu}
    \affiliation{European XFEL, Holzkoppel 4, 22869 Schenefeld, Germany}

	\author{Eugene Bulyak}
	\email{bulyak@kipt.kharkov.ua, eugene.bulyak@desy.de}
	
	\affiliation{National Science Center `Kharkiv Institute of Physics and Technology', 1 Academichna str, Kharkiv, Ukraine\\ V.N.~Karazin National University, 4 Svodody sq., Kharkiv, Ukraine}
%	\collaboration{LUXE Collaboration}
	
	\author{Gianluca Aldo Geloni}
	%\email{gianluca.geloni@xfel.eu}
	\affiliation{European XFEL, Holzkoppel 4, 22869 Schenefeld, Germany}
	
	\date{\today}
	
	\begin{abstract}
		X-ray free-electron lasers (XFELs) utilize high-density and high-energy electron bunches which are well-suited to produce Compton back-scattering radiation. Here we study study interaction of such electron bunches during head-on collision with retroreflected X-ray pulses, emitted by an XFEL. Such collisions allow one to conduct experiments on electron-positron pair production, both through trident and gamma-gamma colliding processes. We discuss cost-effective setups to study such processes, taking advantage of the existing conventional as well as proposed X-ray FEL infrastructure. We estimate parameters of the proposed experiments and compare them with other projects under construction.
	\end{abstract}
	
	%\pacs{41.60.-m, 41.75.Ht}
	
	\maketitle
	
	\section{Introduction}
The main purpose of X-Ray Free-Electron Lasers (XFELs) is to generate extremely brilliant, ultra-short pulses of nearly spatially coherent x-rays with wavelengths of order of 1\,\AA, and to exploit them for scientific experiments in a variety of disciplines spanning physics, chemistry, materials science and biology, \cite{nakatsutsumi2014}.
	
The electron  beams that drive the X-ray FEL process is characterized by  unique  in energy, phase-space density, and monochromaticity and  may be applied for experiments in other areas of physics, where one studies collisions with intense pulses of quasimonochromatic coherent photons. Depending on the photon energy of the impinging photons, we consider two types of experiments: (i) experiments in Physics of particles (more definitive Quantum Chromo-Dynamics, QCD, see \cite{bulyak2023use}, and (ii) experiments In Quantum Electro-Dynamics (QED), discussed in this contribution.

Experiments in high-energy QED require high energy in the electron-photon system. That is, inelastic  electron photon scattering -- electron-positron pair production -- requires that the photon energy in  electron-rest frame  (e.r.f.) exceeds four times the electron rest energy,  i.e. more than  $4m_ec^2$. Moreover,  both  electron bunch and  photon pulse, should be of large density, since the cross section of the process is rather small, much less than one barn.

The photon energy attained so far in the e.r.f. is lower than the triple pair production threshold by an order of magnitude. Even employing a state-of-the-art ultraviolet laser with  $\epsilon = 4.66\,\text{eV}$ requires 112~GeV electrons to attain the threshold, which is currently unavailable. The case for the most powerful infrared lasers is even worse.

We propose to significantly exceed the pair production threshold by employing intense retroreflected X-FEL photon pulses.
	
%	\subsection{Experiments in QED}

At the time of writing, experiments on strong-field QED involving a powerful lasers were carried out \cite{bamber99} and new ones are being designed \cite{bai21, abramowicz21}. 
The most recent paper -- a proposal for Compton-based $\gamma\gamma $ collider intended for the production of  Higgs bosons \cite{barklow23} -- reported a similar idea of employing electron--X-ray photon scattering to produce high-energy gammas.  Also a project to study the storage of multiple X-ray photon pulses  in a cavity-based X-ray FEL and then retro-reflected is now under development \cite{rauer2023}. 

Our proposed arrangement for the interaction the retro-reflected X-ray FEL radiation with the electron bunches 
will  provide (i)  polarized gamma-ray photons in the energy range of about 10~GeV (higher than ever  achieved ), and (ii)  experiments on QED with the energy of gamma-ray photons in the e.r.f. sufficiently exceeding the threshold of the electron-positron pair production.
	
The paper %-- Proposal part II -- 
is organized as follows: after this introduction, section \ref{sec:two} briefly surveys the properties of Compton back-scattered radiation for incident photons with high e.r.f. energy. In the following section we discuss the physics of electron-positron pair generation. Section \ref{sec:four} contains conceptual schemes for electron-photon collisions at X-ray FEL facilities, aiming at  conducting QED experiments  at extremely high energies in the photon-electron system. Finally, the paper is concluded with a summary.       

\section{Interaction of electrons with photons at high energy \label{sec:two}}
Let us consider linear interactions of electrons with photons, i.e. those involving a single electron and a single photon. We focus on high-energy interactions, which are characterized by a high  photon energy in the e.r.f.  $\eta \gtrsim 1$, 
\[
\eta := 2\gamma\omega_\text{las}\; ,
\]
where $\gamma $ is the energy of the electrons in the natural system (coinciding with the Lorentz factor), and $\omega_\text{las} := \epsilon_\text{phot}/m_ec^2$ is the equivalent Lorentz factor for the incident photons. Here and below we assume head-on collisions.  Further we use the natural system of  units for particle and atomic physics, $\hbar = c = m_e = 1$, where appropriate.

\subsection{Compton back-scattering process}
The Compton effect  -- electron-photon elastic scattering --  is equivalent to the classical Thomson process in the limit for low energy of incident photons in the e.r.f., but at higher energies exhibits modification of the spectra and angular distribution of the scattered photons due to the recoil effect.
	
Compton sources of X- and gamma-rays are characterized by two main parameters, namely  (i) the energy of photons and (ii) the intensity of the photon flux.

\subsubsection{Maximum energy in spectrum}
Compared to the case of Thomson scattering, the presence of electron recoil in Compton scattering reduces the energy of scattered-off photons and increases their angular spread.
We account for these effects by introducing an effective electron energy $\gamma_*\le \gamma$. Hence, the energy of a photon scattered  by the electron in the linear  approximation reads,  \cite{DANGELO20001,heinzl13},
	\begin{align}\label{eq:energy}
		&\omega  \approx \frac{4\gamma^2_* \omega_\text{las}}{1+\gamma^2_*\psi^2} \; ; \\
		&\gamma_* := \frac{\gamma}{\sqrt{1+4\gamma\omega_\text{las}}} \;,
		\nonumber
	\end{align}
where $\omega := \epsilon / m_ec^2$ is the energy of the resulting \textit{scattered} photon in the natural system of units (that is the equivalent Lorentz-factor of a scattered photon of energy $\epsilon$), 
$\gamma_*$ is the effective Lorentz-factor of the electron, $\omega_\text{las} := \hbar \omega_\text{las}/ m_e c^2$ is the equivalent Lorentz factor for the \textit{incident}, to-be-scattered, photon of frequency $\omega_\text{las}$, $\psi $ is the scattering angle of the photon relative to the electron trajectory.  We recall the assumption of the head-on collision.
	
The effective Lorentz factor of a `laser photon' in the electron rest frame is, see  \cite{abramowicz21,fedotov23}:
	\begin{equation} \label{eq:eta}
		\eta = 2\gamma \omega_\text{las}\;  .
	\end{equation}

\begin{figure}[htb]
		\includegraphics[width=\columnwidth]{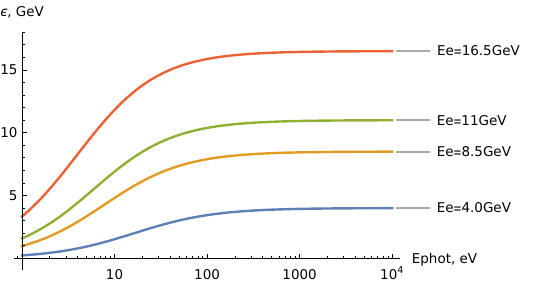}
		\caption{Maximum energy  of back-scattered gamma-ray photons as a function of the energy of incident (`laser') photons. \label{fig:Egvsund}}
	\end{figure}

Under these assumptions, the kinematic relation  \eqref{eq:energy} yields a simple expression for the maximum energy of scattered  photons:
	\begin{equation}
		\omega ^\mathrm{max} = \gamma \frac{4\gamma\omega_\text{las}}{1+4\gamma\omega_\text{las}}\;,
	\end{equation}
which reveals  the recoil effect  governed by the term $4 \gamma \omega_\text{las}$.  For relatively weak electron recoils, $4 \gamma \omega_\text{las}\ll 1$, the maximum energy of the scattered off photons scales linearly with the laser photon energy and quadratically with electron energy:
	\[
	\omega \approx 4 \gamma^2 \omega_\text{las}\, .
	\]
as illustrated in Fig.~\ref{fig:Egvsund}  in the region where $E_\mathrm{phot}<10\,\text{eV}$. 
	
Once the recoil becomes significant,  $4 \gamma \omega_\text{las}\gg 1$,  the energy of  the scattered photons scales linearly with the electron energy and depends only weakly on the energy of the incident photons:
	\[
	\omega = \frac{\gamma}{1+1/(4\gamma \omega_\text{las})}\approx \gamma - \frac{1}{4\omega_\text{las}}\, ,
	\]
as illustrated in the region where  $E_\text{phot}>100\,\text{eV}$ region in Fig.~\ref{fig:Egvsund}. The electrons transfer their almost entire energy to the gamma-ray photons.
Hence, the maximal energy of the backscattered gamma-ray photons asymptotically approaches the electron kinetic energy, while  the \emph{minimal energy} of the recoiled electron drops to
\[
	\gamma'_\text{min} = \frac{\gamma}{1+4 \gamma\omega_\text{las}}\; .
\]
	
	Eq.~\eqref{eq:energy} describes the kinematical dependence of the gamma-ray photon energy on the angle of emission.  In the low-energy Thomson limit, $\gamma_*\approx \gamma$, the energy of gammas drops by half at the emission angle of $1/\gamma $. In the e.r.f. it corresponds to a scenario when a photon scatters perpendicularly to the incident direction.
	Increasing the incident photon energy $\omega_\text{las}$, the half-maximum angle scales as
	\[
	\left(\gamma \psi \right)_{1/2} = \sqrt{1+4\gamma\omega_\text{las}}\; .
	\]

\subsubsection{Spectrum and cross-section}
A high-energy approximation for the Klein-Nishina  formula was derived by Arutyunian and Tumanian \cite{arutyunian63} (see also \cite{hajima2016}). One has the following dependence of the cross section on the frequency of the scattered photons and electron's energy:
	\begin{widetext}
		\begin{align} \label{eq:spectr}
			\frac{\D \sigma_\text{Com} }{\D \omega} &=
			\frac{3 \sigma_\text{Th}}{16\gamma^2 \omega_\text{las}}\left[\frac{\omega^2}{4\gamma^2 \omega_\text{las}^2(\gamma-\omega)^2}-\frac{\omega}{\gamma \omega_\text{las}(\gamma-\omega)}+\frac{\gamma-\omega}{\gamma}+\frac{\gamma}{\gamma-\omega}\right]\mathrm{\Theta}\left(\frac{4\omega_\text{las}\gamma^2}{4\omega_\text{las}\gamma+1}-\omega\right)\; ,\\
			\intertext{while the integral is}
			\sigma_\text{Com} &=
			\frac{3  \sigma_\text{Th}}{32\gamma^3\omega_\text{las}^3}\left\{\frac{4\gamma \omega_\text{las}(2\gamma \omega_\text{las}(\gamma \omega_\text{las}+4)(2\gamma w+1)+1)}{(4\gamma \omega_\text{las}+1)^2}+\left[ 2\gamma \omega_\text{las}\gamma \omega_\text{las}-1)-1\right] \log(4\gamma \omega_\text{las}+1)\right\}\; , \label{eq:comcs}
		\end{align}
where $ \sigma_\text{Th} = 8\pi r_0^2/3$ is the Thomson cross section ($r_0$ the classical electron radius), $\mathrm{\Theta}(\cdot ) $ is the Heaviside Theta function.  
	\end{widetext}
	
\begin{figure} %[htb]
		\includegraphics[width=\columnwidth]{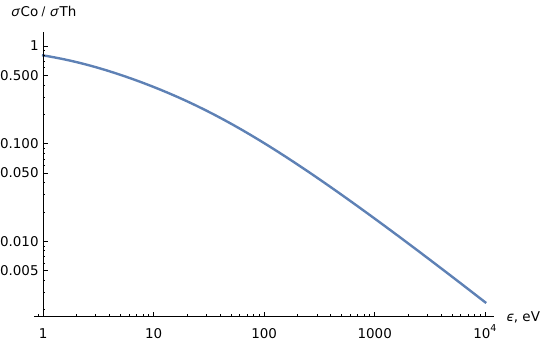}
		\caption{Ratio  of the Compton cross section to the Thomson cross-section for $E_e = 16.5\,\text{GeV}$ vs. energy of incoming photons.  \label{fig:comtho}}
	\end{figure}

The Compton radiation spectrum, given by Eq.~\eqref{eq:spectr}, depends on the energy of  incident photon measured in the e.r.f., $2\gamma\omega_\text{las}$. The total cross section, Eq.~\eqref{eq:comcs}, decreases with the energy of the  incident photon, as shown in Fig.~\ref{fig:comtho}. 

	\begin{figure} %[htb]
		\includegraphics[width=\columnwidth]{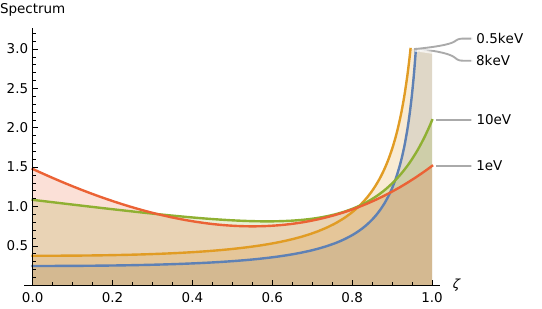}
		\caption{Spectra of Compton radiation, normalized to unity integral, for incoming photon energies of 1~eV, 10~eV, 0.5~keV and 8~keV, $E_e = 16.5\,\text{GeV}$, $\zeta = \omega / \omega_\text{max}$.  \label{fig:specs}}
	\end{figure}
		
\begin{figure} %[htb]
		\includegraphics[width=\columnwidth]{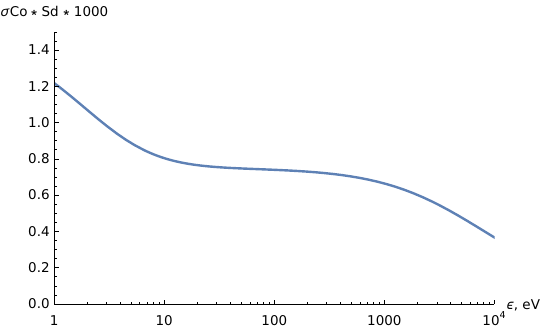}
		\caption{Spectral luminosity of inverse Compton scattered gamma-ray photons within 0.1\%  bandwidth at the high-energy end of the spectrum, calculated as a function of energy of the incident photon energy, assuming an electron beam energy $E_e = 16.5\,\text{GeV}$.  \label{fig:sigsd0001}}
	\end{figure}

	The decrease of the cross-section is accompanied by a change in the spectrum of the back-scattered gamma-ray photon, namely raising and narrowing of its high-energy maximum, as shown in Fig.~\ref{fig:specs} (c.f. \cite{hajima2016}).
	Such redistribution of the spectrum partially compensates the drop of  cross-section, when the peak spectral luminosity is considered:  Fig.~\ref{fig:sigsd0001} presents  the Compton spectral brightness (recall   the Thomson's equal to $1.5\times 10^{-3}$ of total yield).
	
\begin{figure}
	%	\centering
	\includegraphics[width=\columnwidth]{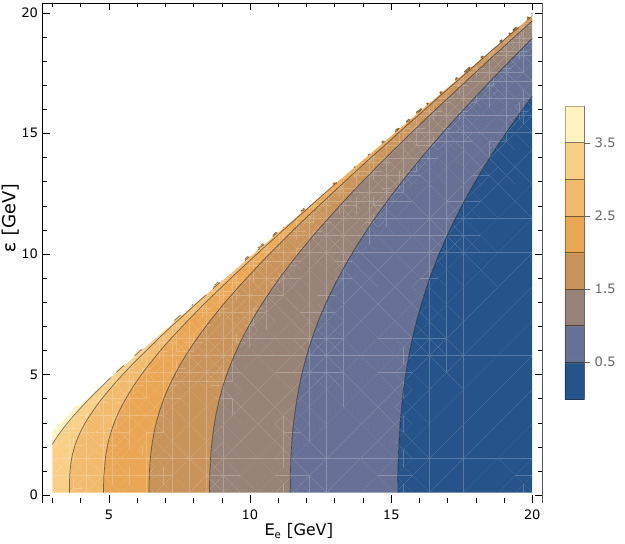}
	\caption{Logarithm of the Compton differential cross section, $\lg (\D\sigma / \D\omega)$ (arb. units), as a function of electron $E_e$ and scattered photon energy $\epsilon$. \label{fig:enpol8kev}}
\end{figure}

The spectral luminosity presented in Fig.~\ref{fig:enpol8kev} is calculated accounting for the quadratic dependence of the FEL photons energy on the electrons energy (in other words we assume a fixed-gap undulator). As it can be seen from Fig.~\ref{fig:enpol8kev}, the spectral luminosity at maximum photon energy exhibits a relatively weak degradation with the increase of the electron energy, while the entire spectrum becomes narrower with its maximum at the maximal energy cutoff. 

The quasi-monochromatic  nature of the Compton radiation at low electron beam energies is actually spoiled out by both the finite energy spread within the beam and, even more significantly, by its finite emittance, see  \cite{bulyak14a}. An increase of the impinging photon energy in the e.r.f., $\epsilon_\text{las} $, allows one to reduce the Compton spectrum bandwidth, as we shown in Table~\ref{tab:fwhh} .  
	
	\begin{table}
		\caption{Spectrum generated by 16.5~GeV electrons \label{tab:fwhh}}
		\begin{ruledtabular}
			\begin{tabular}{lrrr}
				%	\hline
				$\epsilon_\text{las} $, eV  & $\epsilon_\text{max} $, GeV&FWHM,GeV&FWHM / $ \epsilon_\text{max} $\\
				\hline
				$10^0$& 3.33 & 1.31 & $4.0\times 10^{-1}$ \\
				$10^1$& 11.82  & 1.97  &$1.7\times 10^{-1}$  \\
				$10^2$&   15.87&0.53   & $3.4\times 10^{-2}$  \\
				$10^3$&   16.44&  0.062 & $3.8\times 10^{-3}$  \\
				%			8000&   &   & $4.8\times 10^{-4}$  \\
				$10^4$&16.49   &0.0064   & $3.9\times 10^{-4}$  \\
			\end{tabular}
		\end{ruledtabular}
	\end{table}

\section{Electron-positron pair generation}
Besides the generation of  Compton back-scattered gamma-ray photons, the head-on interaction of hard X-rays with high-energy electrons facilitates interesting quantum electro-dynamical processes.

%\subsection{Essential parameters of Quantum Electrodynamics}
Two parameters determine the properties of the photon-electron interactions relevant to Quantum Electrodynamics (QED) (see, e.g. \cite{fedotov23}): (i) the photon energy in the e.r.f.,
\[
\eta := \gamma \omega_\text{las} (1+\cos \theta)\approx 2\gamma \omega_\text{las}\; ,
\]
and (ii) number of the (coherent) laser photons within the Compton wavelength,
\[
\xi :=  \frac{e\mathcal{E}_\text{las}}{m_e\omega_\text{las}}\; ,
\]
where $e, m_e$ are the charge and mass of an electron, resp., $\mathcal{E}_\text{las}$ is the average electric field in the laser pulse.

The first parameter, $\eta $, determines a threshold magnitude: the
photon energy should be sufficient to produce an electron-positron pair, $\eta_* \ge 4$ (explained below).
The second parameter, $\xi $, defines the applicability of the weak-field approximation ($\xi\ll 1$, e.g. perturbative) or the strong-field approximation ($\xi \gtrsim 1$, non-perturbative), see \cite{abramowicz21}.
For the X-rays-on-electron scattering scheme, $\eta $ may reach magnitudes as high $1.2\times 10^3$ at 17.5~GeV.
The other parameter, $\xi $,  is rather low for this scheme, $\xi \approx 2.7\times 10^{-3} $. 
It is worth noting that the maximum energy of laser photons in the e.r.f.  attained so far is $\eta = 0.4$, see the experiment E144 \cite{bamber99} and Figure~2 in \cite{fedotov23}.

\subsection{On the gamma-ray photons energy thresholds}
Let us consider two processes of quantum electrodynamics that can be studied at X-ray FELs: The trident positron production, 
\begin{subequations}\label{eq:epcons}
	\begin{align}
		\gamma + e^-&\to e^-+ e^-+e^+\; ; \label{eq:consa}\\
		\intertext{and the Breit-Wheeler process -- a pair creation from the collision of two photons,}
		\omega +\omega' &\to e^-+e^+\; . \label{eq:consb}
	\end{align}
\end{subequations}
We focus on \emph{the minimal energy} necessary for these processes.

The system of two equations that describes kinematics of the process, i.e., the energy and momentum conservation laws for the process \eqref{eq:consa} reads:
\begin{subequations}\label{eq:mecons}
	\begin{align}
		\gamma +  \omega&= 3\gamma '\; ; \label{eq:energ} \\
		\sqrt{\gamma^2-1}-\omega &= 3\sqrt{\gamma'^2-1}\; . \label{eq:momen}
	\end{align}
\end{subequations}
Here we assume the existence of a frame where all secondary particles, the positron and both electrons, are at rest.

The minimum energy of a photon to produce the electron-positron pair immediately follows from the system \eqref{eq:mecons}:
\begin{equation}\label{eq:wmin3}
	\omega_* = \frac{4}{\gamma+\sqrt{\gamma^2-1}}\; .
\end{equation}
In the e.r.f., $\gamma = 1$, and the minimum photon energy equals four electron masses, $\omega_* = 4$.  The same holds in the ultrarelativistic case, $\gamma\gg 1$: $\eta_* =2\gamma \omega_* \approx 4$.

Therefore, the photon energy should exceed four electron rest energies, $E_\gamma^\text{(r.f.)}\ge 2.044\,\text{MeV}$  to produce the electron-positron pair, see \cite{motz69}. 

\begin{figure} %[htb]
	%	\centering
	\includegraphics[width=\columnwidth]{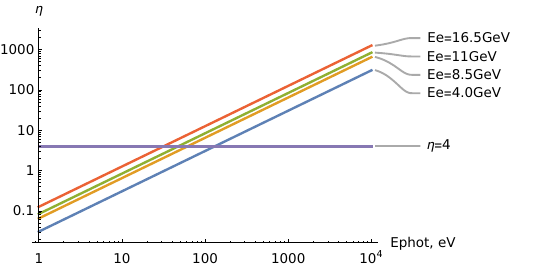}
	\caption{Reduced energy of the photons, $\eta $ in the e.r.f. vs. the photon energy in the laboratory frame. \label{fig:eta2}}
\end{figure}

Figure~\ref{fig:eta2} presents the energy of an incident photon (in  units of the electron rest mass, observable in the e.r.f. ) as a function of the photon energy in the laboratory frame for several electron energies. Evidently, incident optical photons do not suffice for electron-positron pair production at realistic electron beam energies.

\begin{figure} %[htb]
	%\centering
	\includegraphics[width=\columnwidth]{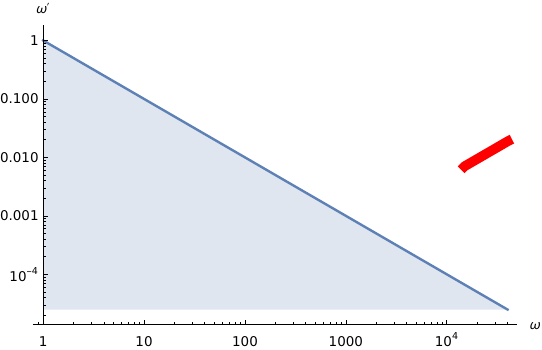}
	\caption{Energy separatrix in the plane $(\omega, \omega')$ . No pair production is possible in the shaded area (We depict the operational range of the European XFEL with a red line).  \label{fig:grest}}
\end{figure}

The same approach may be applied for the `gamma-gamma colliding' scheme (Breit--Wheeler process). Given the conservation of energy and momentum
\begin{subequations}\label{eq:meconsg}
	\begin{align}
		\omega +  \omega'&= 2\gamma \; ; \label{eq:energg} \\
		\omega -  \omega'&= 2\sqrt{\gamma'^2-1}\;  \label{eq:momeng}
	\end{align}
\end{subequations}
and taking $\omega\ge \omega'$ one obtains
\begin{equation}\label{eq:ggsep}
	\omega_*\omega'_*=1\; ;\Rightarrow\; \omega'_*=1/\omega_*\; .
\end{equation}
The energy separatrix on the plane $(\omega,\omega')$ for the Breit-Wheeler process, Eq.~\eqref{eq:ggsep}, yields the condition for the pair production. This criterion is not satisfied in the shaded area on Fig.~\ref{fig:grest}. 

As it can be seen from Eq.~\eqref{eq:ggsep} and Fig.~\ref{fig:grest}, the bigger is the difference between energies, $\omega - \omega'$, the larger becomes the photons total energy, $\omega + \omega'$, necessary for the pair production to take place. For example, 1~eV photons require at least 260~GeV photons to collide with, while 0.511~MeV photons can produce the pairs in collision with photons of equal energy.

\subsection{Cross section  of the trident process}
For the trident process of electron-positron pair production, $e^-+\gamma\to e^-+e^-+e^+$ (pair production by collision with a free electron), the cross section is given by the Borsellino formula for triplet production at very high energies, Eq.~4B-0003 in \cite{motz69}):
\begin{widetext}
	\begin{align}\label{eq:borsel}
		\sigma_\text{triplet} &= \alpha_\text{fs}r_0^2 F(k)\; ; \\
		F(k) &:=
		\frac{28}{9}\log(2k)-\frac{218}{27} - \frac{1}{k}\left[
		\frac{4}{3}\log^3(2k)-3\log^2(2k)+6.84\log(2k)-21.51\right]\; . \label{eq:fk}
	\end{align}
	where $\alpha_\text{fs}$ is the fine structure constant, $r_0$ the classical electron  radius, $k > 100$ the photon momentum in the electron rest frame and in our case $k = 2 \omega \gamma $.
\end{widetext}

\begin{figure} %[h]
	%	\centering
	%	\includegraphics[width=\columnwidth]{Fk.pdf}\\
	\includegraphics[width=\columnwidth]{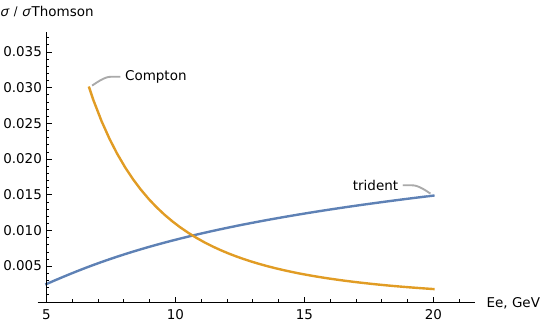}
	\caption{Compton and trident cross sections normalized to  the Thomson's  vs. electron's energy provided that at 16.5~GeV the undulator radiates 8~keV photons. \label{fig:Fk}}
\end{figure}

In Fig.~\ref{fig:Fk} we show the dependence of  Compton and trident cross sections normalized to  the Thomson cross section  (recall that $\sigma_\text{th} = 8\pi r_0^2/3$) as a function of the electron energy, assuming a fixed gap undulator, resonant at 8~keV photon energy for 16.5~GeV electron energy.

As it can be seen from the figure, within the working energy range of the European  XFEL, both the Compton and the trident cross sections are of the same order of magnitude, but their dependence on the electron energy is opposite: the Compton cross-section decreases  while the trident cross-section grows.    

\subsection{Yield of photons/pairs}
The yield of gamma-ray photons (and electron-positron pairs) per single crossing of the incident photon pulse with the electron bunch, both assumed to have 3D Gaussian shapes,  for the head-on collision is independent of the longitudinal dimensions, \cite{bulyak05}. It reads
%	\begin{align*}
	\begin{equation}\label{eq:yield}
		Y = \frac{N_\text{pp}N_e\, \sigma_\text{C,tr} }{2\pi
			\sqrt{\left({\sigma'}_z^2+\sigma_z^2\right)\left(\sigma_x^2+{\sigma'}_x^2 \right)} }\; ,
	\end{equation}
	where $N_\text{pp}, N_e$ are populations of the radiation pulse and the electron bunch, resp., $\sigma_{x,z}, \sigma'_{x,z}$ are their horizontal and vertical dimensions, and $\sigma_\text{C,tr} $ are the Compton/trident cross-section.

	\section{Incident photons: retroreflected hard X-ray self-seeded X-ray FEL radiation \label{sec:four}}
	
High-energy electron-photon interactions at XFEL facilities are an attractive option, because they  may naturally utilize the X-ray FEL radiation emitted by the electron beams and reflected back towards the electron beams, see e.g. \cite{hajima2016}. We exemplify the proposed setup based on existing parameters and components of the self-seeded SASE2 undulator line of the European XFEL, see \cite{sinn12}.
	
The proposed setup implies generation of hard X-ray FEL pulses by the electron beams in an undulator and retroreflecting these radiation pulses 180~deg back towards the counter-propagating electron beams. 
	
\begin{figure*} %[h]
		\includegraphics[width=\textwidth]{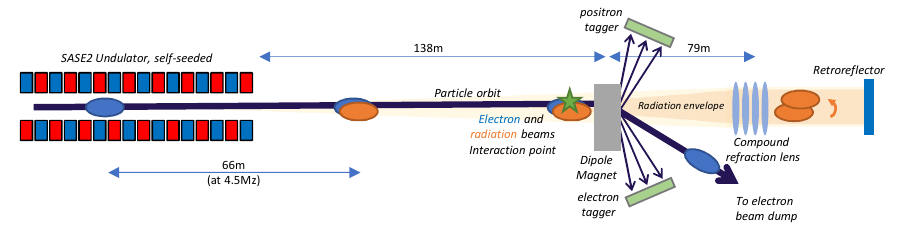}
		\caption{Proposed setup illustration, based on the SASE2 beamline of the European XFEL. The electron beams generate FEL radiation pulses and co-propagate with them until the electrons are deflected. The radiation can be reflected and refocused back on the electron beams. \label{fig:retroreflection_SASE2}}
	\end{figure*}

A sketch of the setup is presented in Fig.~\ref{fig:retroreflection_SASE2}. The highest intra-train repetition rate of the European XFEL accelerator is 4.5~MHz. This corresponds to a minimum separation between the electron bunches as well as the emitted FEL radiation pulses of 66~m. After being emitted, the radiation pulses co-propagate with the electron beam for another 138~m before being finally separated, conceptually allowing installation of diagnostics equipment.
	
As long as the FEL is tuned to generate hard X-ray radiation, the retro-reflection can be carried out with crystal optics. The reflector can be comprised of a single crystal  satisfying the Bragg condition at 90 deg incidence, or of two crystals fulfilling the Bragg condition at 45~deg. Diamond, Silicon or Germanium can serve as  crystal materials. To efficiently reflect the FEL radiation, the spectral bandwidth of the reflector should ideally fit the bandwidth of the FEL radiation. To maximize the reflection efficiency, we both decrease the radiation bandwidth and increase the spectral acceptance of the reflector.
	
A single diamond crystal Hard X-ray self-seeding method \cite{geloni11} has demonstrated mono\-chromatization of FEL radiation at the European XFEL, yielding 1\,mJ pulses with 1.3\,eV fwhm bandwidth at 7.5\,keV, and 0.8\,mJ within 0.7\,eV bandwidth at 13\,keV photon energy\,\cite{Liu2023}.
	
A diamond retroreflector offers a rather narrow reflection bandwidth of about 0.1 eV FWHM, allowing to reflect only about 1/10th of the initial pulse energy of the self-seeded FEL radiation. However, this is compensated by diamond higher resistance to radiation damage over the MHz-rate pulse train, allowing to utilize the full repetition rate of the European XFEL beamline of thousands pulses per second (theoretically up to 27000 pulses per second).
	
\begin{figure}
		%	\centering
		\includegraphics[width=\columnwidth]{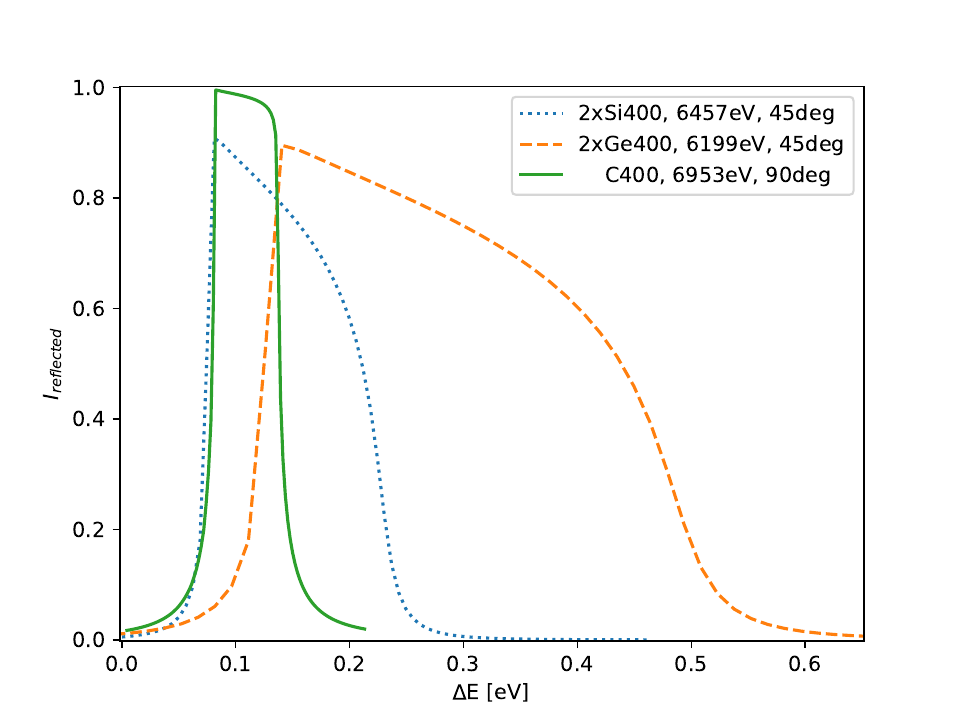}
		\caption{Reflectivity curve of a retroreflector comprising different crystals in a single- (90~deg incidence angle) and two-crystal (45~deg incidence angle on each crystal) geometries. The Bragg condition determines  photon energies suitable for different designs.\label{fig:reflections}}
	\end{figure}

At low repetition rate accelerators one can employ a 2-bounce reflector with Silicon or Germanium crystals with (100) cut and (400) reflections, as the large interplanar distance and low Bragg angles allows to maximize the reflector bandwidth and thus also the amount of reflected photons. Here material and geometry determine a target photon energy of the setup of about 6457 or 6199~eV for Silicon and Germanium, yielding a reflection bandwidth of 0.2 and 0.4~eV respectively. The reflectivity curves of different retroreflector configurations were calculated with the XOP software package \cite{xop1997} and presented in Fig.~\ref{fig:reflections}.  
	
To increase photon density at the interaction point, i.e. to reduce $\sigma_{x,y}$, we suggest collimating the radiation with Compound Refractive Lenses (CRLs) before the retro-reflection. This would both minimize the angular spread of the radiation impinging on the retroreflector in order to satisfy the Bragg condition, as well as to re-image the radiation on its return path. Given the flexibility of the CRL focal length setup at the SASE2 beamline of the European XFEL (see Section~4.1 of \cite{nakatsutsumi2014}), such collimation would allow one to re-image the radiation pulses back on an electron beam at the desired interaction point.
	
To satisfy the simultaneous arrival of both electron and photon beams to the interaction point, the latter should be located at a multiple of 33~m from the retroreflector.
The electron- (hence also the photon-) beam arrival time jitter at the European XFEL is below 30~fs, \cite{Sato2020}. This results in a longitudinal jitter of the interaction point by less than hundred micrometers facilitating a precise location of the interaction point.

Upon  re-imaging, the radiation waist can be geometrically demagnified compared to the source in the undulator by about a factor 1.5, depending on its position. This can be coupled with a modified electron optics to reduce the transverse size of the electron beam as well. Reducing either contributes to higher yield of the  electron-photon interactions.
	
	Given the European XFEL performance and the beamline geometry we estimate the interaction of a 50\,$\mu$J, 7\,keV radiation pulse and 20\,$\mu$m FWHM transverse size with a 14\,GeV electron beam of comparable size.
	
	A \textit{single interaction} of photon- and electron beams would result, on average in: $10^{-2}$ gamma-ray photons, $10^{-2}$ sub-GeV electrons, $10^{-2}$ particles with 1/3 of the electron beam energy generated via the trident process, a rare pair of particles with half the initial electron beam energy generated via the Breit-Wheeler process.
	
	We estimate the yield of positrons with Eq.~\eqref{eq:yield} by substituting the trident cross section for the Compton cross-section.  Detection of electrons and positrons would ideally require low-noise single-particle counting detectors located downstream the interaction point. 
	%For proof-of-principle experiment an existing infrastructure may potentially be used, e.g. electron energy spectrometer\,cite{tomin2022}.

	\subsection{Employing the radiation stored in cavity-based XFELs}
	
	Another scheme allowing electron-photon collisions with parameters comparable to those outlined in the previous Section utilizes the radiation stored in the resonator of an XFEL Oscillator (XFELO). A notable benefit of this proposal is its substantial added value to otherwise unperturbed generation of narrow-bandwidth X-ray FEL radiation at an XFELO setup. 
	
\begin{figure*} %[h]
		%	\centering
		\includegraphics[width=\textwidth]{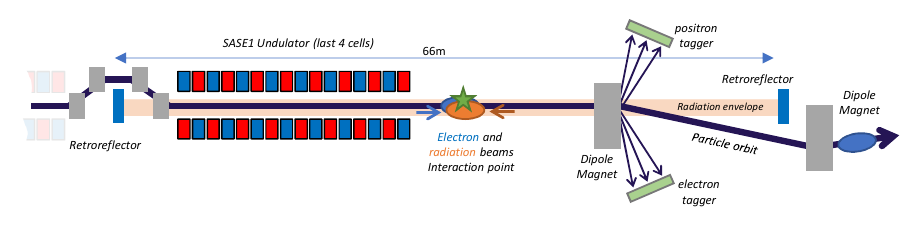}
		\caption{Illustration of the proposed setup, based on XFELO proposal at SASE1 beamline of the European XFEL. The electron beams generate FEL radiation pulses and co-propagate with them until the electrons are deflected to avoid hitting a retroreflector. The latter reflects the radiation back on the electron beams. \label{fig:retroreflection_SASE1}}
	\end{figure*}

	We exemplify this scheme on the slightly modified proposed setup for demonstrating cavity-based Free-electron laser at the SASE1 undulator of the European XFEL \cite{rauer2022, rauer2023}. This setup, illustrated on Fig.~\ref{fig:retroreflection_SASE1}, employs a set of two C400 diamond retroreflectors comprising a 66\,m-long oscillator cavity, resonant to 6.95~keV, with the radiation amplification taking place in an undulator via  FEL interaction. The radiation pulse energy in the cavity is expected to increase up to a 10~mJ or 10~$\mu$J-order level, depending on the implemented hardware.
	
	As the designed distance between the retroreflectors is expected to be 66~meters, the reflected X-ray radiation will meet the following electron bunch exactly in the middle of the cavity provided 4.5~MHz repetition rate of the accelerator. Upon minor modification of the oscillator geometry, the electron-photon interaction would take place downstream the last cell of  the SASE1 undulator. 
	
	The particle yield per interaction would be comparable to that discussed in the previous section. Given the non-invasive, parasitic nature of this method, aside of the need for additional electron energy diagnostics, it may be considered as a cost-effective upgrade to  planned cavity-based project at the European XFEL and, more in general, at other X-ray FEL facilities worldwide.
	
	Regardless of the method, retroreflecting the X-ray FEL radiation towards the counter-propagating electron beam would allow one to generate monochromatic Compton-scattered gamma-ray quanta with energies nearly equal to that of the electron beam.
	
	\section{Summary and Discussion}

	We proposed a scheme, based on the collision of retroreflected, quasimonochromatic X-ray FEL radiation pulses with the high-energy and dense electron bunches in the same FEL. We exemplified the scheme with two different generation and retro-reflection methods at two beamlines of the European XFEL facility, taking into account its geometry, radiation and electron beam parameters, as well as  already-proposed upgrades.
	
%	We conclude that in addition to producing the extremely bright X-rays, the XFEL accelerator allows to conduct experiments in chromodynamics with polarized photons in the energy range of  2.5--4~GeV. This range would cover the gap between LEPS2 and GlueX gamma-ray beamlines currently under construction.
	
	Successful retroreflection of X-ray FEL radiation will allow extending  the energy range of the generated gamma-ray photons up to 10--15~GeV -- the highest among  proposed designs. In addition, the gamma-ray radiation, generated by scattering multi keV photons on multi GeV electrons, will be almost monochromatic (relative bandwidth circa $5 \times 10^{-4}$, and would therefore not require coincidence schemes for photon energy measurement.
	
	The energy of  retroreflected photons in the electron rest frame will sufficiently exceed the threshold of both the trident and Breit-Wheeler processes of positron-electron pair production, providing  conditions for experiments in QED. 

		Upon collision between the X-ray FEL radiation pulse and the electron bunch  a \emph{unique electro-dynamical entity} will be generated, composed of the relativistic electrons and positrons along with X-ray- and gamma-ray photons: it will be the subject for further studies. The composition and roughly estimated parameters of the collision remnants are listed in Table~\ref{tab:elsum}.
	
	\begin{table*}
		\caption{Electrodynamics. Summary table \label{tab:elsum}}
		\begin{ruledtabular}
			\begin{tabular}{lllr}
				particles & num/bunch & energy & notes\\
				\hline
				initial bunch& $N_e\approx 10^{10}$ & $E_e = 4-16.5$, GeV & (0.25--1)\,nCoul \\
				x-ray photons&$N_x\sim 10^3 N_e$    & $E_x = (1-8)$\,keV &   \\
				Compton photons& $N_\gamma\sim 10^{-11} N_e$  & $E_\gamma = (2-16.5)$ GeV &    \\
				pr. el-pos pairs via trident& $N_{+,-}\lesssim 0.02$  & $ E_{+,-}\approx E_e / 3$ & $e^-\gamma\to e^-e^-e^+$  \\
				sec. el-pos pairs via Breit-Wheeler &    $N'_{+,-} < N_\gamma$   &$E_{+,-}\approx E_\gamma / 2$&$\gamma\gamma'\to e^-e^+$   \\
			\end{tabular}
		\end{ruledtabular}	
	\end{table*}

In such composite bunches, the physical processes, e.g., electron-positron annihilation, proceed $\gamma $ times slower than in the case of non-relativistic speed. The symmetric magnet spectrometer -- as depicted in Figs.~\ref{fig:retroreflection_SASE2} and \ref{fig:retroreflection_SASE1}  -- would allow detecting their components: 
	\begin{itemize}
		\item low-energy electrons  (optionally) in coincidence with Compton-backscattered high-energy gamma-ray photons (polarised);
		\item electrons and positrons, in coincidence, each of energy $E_e/3$, revealing the $e^-\gamma\to e^-e^-e^+$ reaction;
		\item electron-positron pairs, in coincidence, with energy $\approx E_e/2 $ indicating the $\gamma\gamma'\to e^-e^+ $ reaction.
	\end{itemize}

	It is worth to emphasize the dominance of the trident process over the Compton scattering and the Breit-Wheeler process in collisions of the X-ray photons with GeV-energy electrons.

	From a ``black box'' point of view, there is no difference between the trident and the two-step (Compton + Breit-Wheeler) processes of positron creation: the relativistic electron enters the interaction region (``black box'') containing nothing but radiation, and the output consists of two electrons and a positron with energies that depend on the process. However, different kinematics of the processes as well as the dominance of the trident process by a few orders of magnitude makes it possible to study this process in detail.
	
	Such experiments, having been considered nowhere else, are characterized by low background, notwithstanding the geometry of  quasimonochromatic beams of electrons and photons colliding head-on.
	
	The proposed setup may be a stepping stone as a lower light intensity stage of the LUXE project \cite{abramowicz21} with similar motivations and goals. Also the setup makes it possible to conduct unique experiments to study the linear (involving two-photon collisions) Breit-Wheeler process and the linear trident process on free electrons.

 The proposed experimental setup will be at the edge of QED experiments: comparable projects -- Ref.~\cite{hajima2016} 
%	Most recent paper
 or Proposal for Compton-based $\gamma\gamma $ collider intended for production of the Higgs bosons \cite{barklow23} -- reported a similar idea of employing electron--X-ray photon scattering to produce high-energy gammas. The former is intended to produce gammas with smaller energy, $\lesssim 7\,\text{GeV}$, the latter is    far more complicated and expensive.

 In addition, the polarized positrons produced in the GeV energy range and with a flux of $1\dots 10$ per second may conceptually be utilized by further experiments in particle physics. 
	
	\begin{acknowledgments}
		The authors would like to thank Patrick Rauer, Harald Sinn, Frederik Wolff-Fabris, Anders Madsen, Jan Gr\"unert, Sergey Tomin for fruitful discussions and Serguei Molodtsov for his interest and support. Computations were done with the \textsc{Wolfram Mathematica Cloud} package \cite{Mathematica}, provided to one of coauthors (EB).
	\end{acknowledgments}

\providecommand{\noopsort}[1]{}\providecommand{\singleletter}[1]{#1}%

\end{document}